# Berry Phase and Anomalous Transport of the Composite Fermions at the Half-Filled Landau Level

W. Pan[1,*], W. Kang[2,*], K.W. Baldwin[3], K.W. West[3], L.N. Pfeiffer[3], and D.C. Tsui[3]

[1]Sandia National Laboratories, Albuquerque, NM 87185 USA

[2]Department of Physics, University of Chicago, Chicago, IL 60637 USA

[3]Department of Electrical Engineering, Princeton University, Princeton, NJ 08544 USA

**The fractional quantum Hall effect (FQHE)[1,2] in two-dimensional electron system is an exotic, superfluid-like matter with an emergent topological order. From the consideration of Aharonov-Bohm interaction of electrons and magnetic field, the ground state of a half-filled lowest Landau level is mathematically transformed to a Fermi sea of composite objects of electrons bound to two flux quanta, termed composite fermions (CFs)[3-5]. A strong support for the CF theories comes from experimental confirmation of the predicted Fermi surface at $\nu = 1/2$ (where $\nu$ is the Landau level filling factor) from the detection of the Fermi wave vector in the semi-classical geometrical resonance experiments[2,6-9]. Recent developments in the theory of CFs[10-21] have led to a prediction of a $\pi$ Berry phase for the CF circling around the Fermi surface at half-filling[10,14,17-20]. In this paper we provide the first experimental evidence for the detection of the Berry phase of CFs in the fractional quantum Hall effect. Our measurements of the Shubnikov-de Haas oscillations of CFs as a function carrier density at a fixed magnetic field provide a strong support for an existence of a $\pi$ Berry phase at $\nu = 1/2$. We also discover that the conductivity of composite fermions at $\nu = 1/2$ displays an anomalous linear density dependence, whose origin remains mysterious yet tantalizing.**

Under the framework of the CF theory[3-5], the effective magnetic field, $B_{eff}$, that the CFs experience in the lowest Landau level is reduced due to flux attachment so that $B_{eff} = B - B_{\nu=1/2}$. At the half filling of Landau level ($\nu = 1/2$) $B_{eff}$ becomes zero and CFs form a Fermi sea state. This startling prediction of the CF theory has been verified experimentally[2,6-9]. Away from $\nu = 1/2$, the $\nu = n/(2n + 1)$, where $n = \pm1, \pm2, \pm3\ldots$, FQHE states (also called the Jain sequence of FQHE states) can be mapped into the $\nu^* = n$ integer quantum Hall effect (IQHE) states of CFs (Fig. S1). The experimentally observed sequence of FQHE states can be viewed as the density of states oscillations of the pseudo-Landau levels of CFs under increasing $B_{eff}$[3-5]. The CF theory has shown to be extremely proficient in constructing the wave functions of the various FQHE states and in providing explanations of the experiments on the FQHE[2,4].

There have been new, interesting developments in the study of CF in the recent years. The interest in CF was reinvigorated in part from the realization of the significance of particle-hole symmetry[10,15-18,20-23]. In particular a recent experiment on the commensurability oscillations of CFs observed that the oscillation minima above and below half-filling are not symmetric with changing magnetic field[23]. This result hinted that the nature of the composite fermions may change from "particle-like" below to "hole-like" above half-filling. While particle-hole symmetry at half-filling had been accepted without any question, its experimental consequence on the CFs at half-filling remained largely unexplored.

Many of theoretical studies of the particle-hole symmetry in the FQHE predict a Berry phase of the composite Fermi liquid at $\nu = 1/2$ [10,17-20]. In the Dirac theory of CFs at finite density[10-14], the CFs at half-filling are treated as Dirac fermions with a $\pi$ Berry phase from its motion about the Fermi surface. An alternate approach, which considers CF as a charge-neutral particle carrying vorticity[17,18], associates a Berry curvature about the CF Fermi surface at the half filling. A microscopic approach to CFs based on geometrical considerations also predicts a Berry phase for adiabatic transport of CFs around the Fermi surface[19, 24]. The connection between the Dirac theory[10-14] and the Halperin, Lee, Read (HLR) theory[3] of CFs was explored recently[20]. These theories provide a various perspectives into the physics of CFs at the half-filled Landau level. The prediction of a Berry phase for CFs in the FQHE is a significant new development as its possibility has been overlooked in the past.

In this paper we study the composite fermions at the half-filled Landau level by studying their density-dependent magnetotransport. To look for signatures of the predicted Berry phase, we have measured the density-dependent Shubnikov-de Haas oscillation of magnetoresistivity, $\rho_{xx}$, around $\nu = 1/2$ at various magnetic fields in two heterojunction insulated-gate field-effect transistors (HIGFETs)[25], in which the electron density ($n_e$) can be tuned over a large range. [See details of the characterizations in Fig. S2 in the Supplementary Information (SI).] Study of SdH oscillations in the FQHE has been a very useful probe of CFs in determining their effective mass in the lowest Landau level[26, 27]. In case of graphene, studies of SdH oscillations have been successfully utilized to detect the Berry phase of Dirac fermions about the Dirac cone[28, 29]. Our Berry phase results can be understood within the framework of the recently proposed theoretical studies of CFs at the half-filled Landau level[10-21].

Fig. 1a shows the SdH oscillations from specimen HIGFET-A as a function of $n_e$ under a fixed magnetic field at B = 10T. A series of well-defined FQHE states is observable as a function of density as SdH oscillation minima around $\nu = 1/2$. Under a fixed magnetic field B, the density-dependent SdH oscillation of the magnetoresistivity, $\Delta\rho_{xx}$ can be modeled as:

$$\Delta \rho_{xx} = R(n_e, T) \cos\left[2\pi\left(\frac{n_B}{n_{1/2} - n_e} + \beta\right)\right] \tag{1}$$

where $R(n_e,T)$ is the amplitude of the SdH oscillations as a function of density and temperature, $n_{1/2}$ is the density at $\nu = 1/2$ for the given B, $n_B$ the frequency of the density dependent SdH oscillations, and $\beta$ the associated Berry phase of the CF motion about the Fermi surface. In our analysis we first locate the densities corresponding to the FQHE states at the magnetoresistivity minima of the SdH oscillations. (The $\rho_{xx}$ maxima can be neglected since only the FQHE states at the minima are the relevant for CFs.) The value of index *n* for the respective SdH minima can be independently determined from the one-to-one correspondence for the Jain states between the IQHE around B = 0 and the FQHE around $\nu = 1/2$. The procedure is illustrated in Fig. S1 in the Supplementary Information. As shown in Fig. S1, after the $R_{xx}$ curve is horizontally shifted so that the magnetic field position at $\nu = 1/2$ overlaps with B = 0, we observe that the $\nu = 1/3$ and 1 states, the 2/5 and 2 states, the 3/7 and 3 states, etc. respectively occur at the same positions. Based on this observation, one can assign an effective integer filling factor to a FQHE state, *i.e.*, filling factor 1 to the 1/3 state, 2 to 2/5, 3 to 3/7, and -2 to 2/3, -3 to 3/5, -4 to 4/7, etc. The effective magnetic field $B_n^* = B - B_{1/2}(n_e)$ where $B_{1/2}(n_e) = 2n_e h/e$ is the magnetic field at half filling for the electron density of $n_e$. The inset of Fig. 1b illustrates how $B_n^*$ is determined in such a density sweep. The values of $B_n^*$ is shown as the top axis in Fig.1a. In Fig. 1b, we plot the SdH oscillation index *n* vs. $1/B_n^*$. An intercept of $\beta = -1/2$ at $1/B_n^* = 0$ is clearly seen in Fig. 1b. This result shows that Berry-phase of $\pi$ for the CFs at $\nu = 1/2$.

We want to point out that similar measurement has been carried out in graphene to characterize its Berry phase[28, 29]. The chiral nature of massless Dirac fermions in graphene produces a Berry phase of $\pi$ about its Dirac point, resulting in SdH oscillations that are phase shifted by π relative to 2D systems with conventional Fermi surface. Experiments have shown that β determined from the intercept is -1/2 in graphene. Our measurement likewise illustrates that the density dependent SdH oscillations of the CFs demonstrate that β = -1/2, which corresponds to a π Berry phase about the Fermi surface at ν=1/2. We note here that in the SdH measurements at fixed electron density, the intercept at $1/B_n = 0$ is zero, yielding a zero for the Berry phase. Here $B_n$ is defined as $B_n = B_\nu - B_{1/2}$ for the effective magnetic field that CFs see in standard SdH oscillations where the external B field is varied. We have shown this in one of our high mobility quantum well samples in Fig. S4 in the SI.

We want to point out that in the proposals of Dirac composite fermions at finite density[10-14] and CF as neutral particle carrying vorticity[17,18], the CFs density, $n_v$, is proportional to the external magnetic field, $n_v = eB/2h$. Only at ν = 1/2 does $n_v$ equal the underlying electron density ($n_e$). In the HLR theory of CFs, the density of composite fermions equals the electron density. It was pointed out by Wang and Senthil that in conventional SdH measurements at fixed density, both the CF density and its effective magnetic field change with the varying external magnetic field[18]. Thus it is not possible to obtain a coherent Berry phase and β becomes zero in SdH measurements at fixed density.

Fig.2 summarizes the measurement and analysis of density-dependent SdH oscillations at different values of fixed magnetic fields. Fig. 2a-c shows SdH oscillations about the ν = 1/2 for three different magnetic fields. Fig. 2d shows the resulting SdH fan diagram of $n$ vs $1/B_n^*$ for the density sweeps in Fig. 2a-c where all the data converge at the vertical intercept of β = -1/2. A summary of slope intercepts β determined from the various density sweeps is shown in Fig. 2e.

The slope-intercept from the density sweeps with different magnetic fields all universally possess a β = -1/2, confirming the predicted Berry-phase of π for the CFs at ν = 1/2

We now turn our attention to the ν = 3/2 state, the particle-hole conjugate state of the 1/2 state. Around ν=3/2, the FQHE occurs at $\nu = 2 - n/(2n+1)$. We measure the SdH oscillations as a function of electron density at a fixed magnetic field and apply a formula of $B_n^* = 3\times(B - 2h/3e \times n_e)$ for the ν=3/2 CFs. With this definition of effective magnetic field, we arrive at $n = -B/2 \times 1/B_n^* - 1/2$, the same as the ν=1/2 CFs. We note here that the minus sign in front of B/2 reflects the hole nature of the CFs at ν=3/2. Fig. 3a shows the SdH oscillations around ν=3/2 as a function of electron density at a fixed magnetic field of 5.0T. The arrows mark a few represented FQHE states. In Fig. 3b, n versus $1/B_n^*$ is plotted. An intercept of -1/2 at $1/B_n^* = 0$ is clearly seen, now establishing a direct detection of a π-Berry phase for the CFs at ν = 3/2. A summary of the slope intercepts determined from the various density sweeps with different magnetic fields is shown in Fig. S5 in the SI. They all universally possess the value of -1/2.

For CFs around ν=3/2, in standard, B-dependent SdH oscillations, the effective magnetic field of CFs around ν = 3/2 is given by $B_n = 3\times(B_\nu - B_{3/2}) = -n_e h/e \times 1/(3n+2)$, or $n = -n_e h/3e \times 1/B_n - 2/3$. The intercept of -2/3 was indeed observed in the standard SdH oscillations, as shown in Fig. S6 in the SI.

In our density dependent SdH measurement, the density of CF depends on the magnetic field at ν = 1/2. The CF density, $n_\nu = eB/2h$, is proportional to the external magnetic field, and only at ν = 1/2 equals to the underlying electron density $n_e$. Our experimental detection of the Berry phase in the density dependent SdH measurement appears to support the proposal that the density of CF is proportional to $B_{1/2}$, the magnetic field at ν = 1/2 and not the electron density. However, our observation of the Berry phase cannot be viewed as an evidence of the Dirac fermions[10-14] or the neutral vortex theories of CF[17,18]. The Berry phase at $\nu = 1/2$ appears to be an emergent, universal feature of the underlying CF liquid[10-24] as electrons seek to minimize their interaction energy in the lowest Landau level.

In contrast to similar experiments in graphene, a Dirac point in the lowest Landau level cannot be directly observed in GaAs heterostructure or similar systems. In the context of the Dirac fermion theory of CFs[10-14], an effective Dirac point may be discussed as an extrapolation from higher lying pseudo-CF Landau levels. A direct detection of such a Dirac point is beyond the scope of the present experiment. It follows that our observation of the $\pi$ Berry phase of CFs establishes an important, previously unrealized correlation of CFs.

In Fig. 4 we present the first ever result of the density dependence of CF conductivity which in principle provides an important insight into the transport of CFs at $\nu = 1/2$. A plot of $1/\rho_{xx}$ at $\nu = 1/2$, obtained in the two HIGFETs, is shown as a function of electron density $n_e$. Since the conductivity of CFs, $\sigma_{CF} \cong 1/\rho_{xx} \times (1+ (\rho_{xx}/\rho_{xy})^2)$ [17] at $\nu = 1/2$, and with $(\rho_{xx}/\rho_{xy})^2 < 0.1\%$, $\sigma_{CF} \cong 1/\rho_{xx}$. Both HIGFETs show that the conductivity of CFs shows a linear dependence over a large density range from ~ $2\times10^{10}$ to $1.4\times10^{11}$ $cm^{-2}$. A linear dependence of conductivity was also observed for the CFs at $\nu=3/2$, as shown in Fig. S7. Differing from the result at $\nu = 1/2$ and 3/2, the conductivity of electrons at $B = 0$ follows a $n_e^2$ density dependence (details shown in Fig. S8 in the SI).

At this point we do not understand the origin of this striking linear density dependence of conductivity at $\nu = 1/2$. While a 2D electron systems in a heterostructure with an appropriate form of disorder may have a conductivity $\sigma_{xx} \sim n^\alpha$, $\alpha \approx 1$ at $\nu = 1/2$[30], it is unclear how well this theory applies to our HIGFET device which has no doping and instead relies on electrostatic gating.

The observed linear density dependence of conductivity at $\nu = 1/2$ may be viewed in terms of a constant CF mobility, $\mu^*$, with $\sigma_{CF} = n_\nu e \mu^*$. Consequently, the scattering time $\tau^* = \mu^* m^*/e \propto n_\nu^{1/2}$ since the effective mass of CFs, $m^* \propto n_\nu^{1/2}$.[31] This apparent $\tau^* \propto n_\nu^{1/2}$ behavior differs from the HLR theory[3] where the CF scattering time $\tau^* \propto m^* k_F \propto n_\nu$, since the Fermi wave vector for CFs, $k_F \propto n_\nu^{1/2}$. Moreover, from the Dingle analysis of the SdH oscillations around $\nu=1/2$ (Fig. S9), it was observed that CF Landau level broadening (or momentum scattering time) is weakly density dependent. More studies will be needed to understand the origin of this weak density dependent behaviour and whether it is responsible for the linear density dependence of CF conductivity.

In the context of the Dirac theory of CFs[10-14] or the neutral vorticity theory[18,19], the density of CF at $\nu = 1/2$ is $n_\nu = eB_{1/2}/2h$. Then the conductivity at the half-filling, $\sigma_{xx} \sim n_\nu \sim B_{1/2}$, provided that $\mu^*$ is constant. Interestingly, it was shown experimentally in ultra-clean specimen that longitudinal resistivity at even denominator fillings are linear with magnetic field[32, 33]. The observed behavior is consistent with our results since $\rho_{xx} \sim \sigma_{xx} \sim B_{1/2}$. Our finding therefore hints at the possibility of some novel exotic entanglement of the CFs at the half-filling.

It is well-known that the conductivity of the 2D Dirac fermions, for example, in graphene displays a linear density dependence, due to its linear dispersion[28,29]. Since there is no experimental evidence yet available for the existence of Dirac fermions at $\nu = 1/2$, it is unclear if there is any connection between the linear density dependence of graphene and the CFs at $\nu =$ 1/2 at this time. At minimum a further clarification of the scattering time will be necessary to better understand this puzzling behavior of the CFs at the half-filled Landau level.

**Methods**: The heterojunction insulated-gated field-effect transistors (HIGFETs)[25] were exploited for the SdH oscillations and composite fermion conductivity measurements. The growth structure of a typical HIGFET and the density dependence of mobility can be found in Ref. [25]. Low frequency (~ 11 $Hz$) lock-in (Princeton Applied Research 124A) technique was used to collect $R_{xx}$ and $R_{xy}$ as a function of electron density by sweeping the gate voltage at a fixed magnetic (B) field. $\rho_{xx}$ is obtained from the measured $R_{xx}$ by taken into account the geometric ratio, and $\rho_{xy} = R_{xy}$ in two-dimensions.

Some of the unique, relevant aspects of HIGFETs are worth pointing out. In a HIGFET device, a heavily doped GaAs top layer serves as the top gate, which is a significant improvement over commonly used Ti/Au or Cr/Au gate for realizing a uniform two-dimensional electron system. A differential thermal expansion between Ti/Au (or Cr/Au) and GaAs is known to induce severe electron density ($n_e$) inhomogeneity when the specimen is cooled from room temperature down to cryogenic temperatures. In addition, the insulating AlGaAs buffer layer in HIGFET devices is MBE grown, considerably more uniform than a dielectric layer commonly used in a field-effect transistor. The density of charge traps is reduced to the lowest level possible, and the linear relationship of $n_e$ versus $V_g$ can hold down to very low density.

The spatial uniformity of electron density can be attested by an observation of a perfectly linear relationship in Figure S2d. Achieving a very low charge traps density in our HIGFET was confirmed by observing a perfect overlap between the two traces of $V_g$ sweeping up and down.

Finally, in all the HIGFET devices we studied, an electron mobility of $\mu \sim 1\times10^6$ $cm^2/Vs$ was achieved at $n_e \sim 2\times10^{10}$ $cm^{-2}$, as shown in Fig. S3. Such a high sample quality effectively eliminates the disorder effects.

**Acknowledgements**: We thank D. Son and F.D.M. Haldane for alerting us of their theories and for invaluable discussions. We also thank J. Jain, N.P. Ong, R. Bhatt, M. Shayegan, D. Feldman, M. Levin, A. Stern, C. Kane, Ashvin Viswanath, P. Zucker, S. Simon and M. Zudov for helpful discussions. This work was supported by the U.S. Department of Energy, Office of Science, Basic Energy Sciences, Materials Sciences and Engineering Division. Sandia National Laboratories is a multi-program laboratory managed and operated by Sandia Corporation, a wholly owned subsidiary of Lockheed Martin Corporation, for the U.S. Department of Energy's National Nuclear Security Administration under contract DE-AC04-94AL85000. The work at University of Chicago was supported in part by the Templeton Foundation and NSF MRSEC Program through the University of Chicago Materials Center. Sample growth at Princeton was funded by the Gordon and Betty Moore Foundation through the EPiQS initiative GBMF4420, and by the National Science Foundation MRSEC Grant DMR-1420541.

**Figures and Figure Captions:**

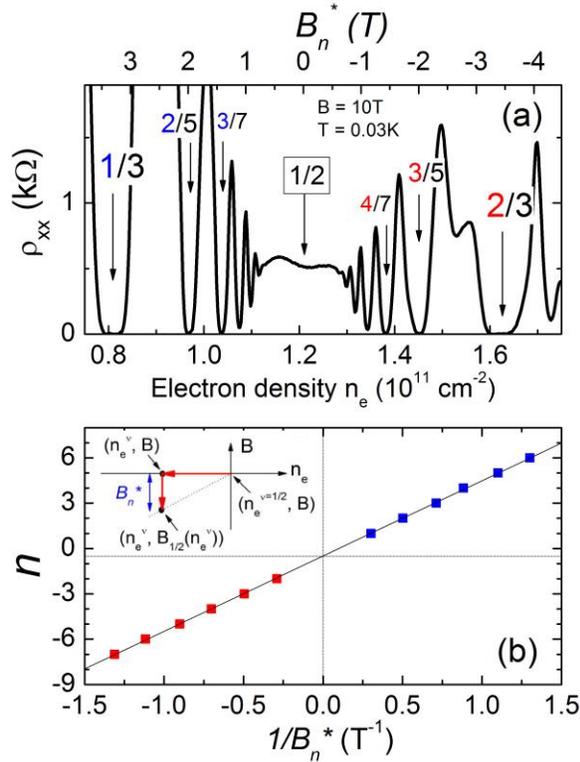

Figure 1: Composite fermion Shubnikov-de Haas oscillations and Landau fan diagram in a heterojunction insulated-gate field- effect transistor (HIGFET). (a) shows $\rho_{xx}$ as a function of electron density ($n_e$, in units of $10^{11}$ $cm^{-2}$). The arrows mark several representative FQHE states. Their corresponding CF pseudo-Landau level filling factor *n* is indicated with the colored numerator, blue for positive and red for negative integers. The index n is determined independently from the one-to-one correspondence between the FQHE around $\nu = 1/2$ and the IQHE around B = 0, as shown in Fig. S1. For every state with index *n*, the corresponding filling factor under the composite fermion theory is given by $\nu = n/(2n + 1)$. The upper x-axis shows the value of $B_n^* = B - 2n_e h/e$. We note that the particle-hole conjugate states, for example the 1/3 (*n*=1) and 2/3 (*n*=-2) states are located equal densities away from $\nu = 1/2$. (b) displays CF pseudo-Landau level filling factor *n* vs $1/B_n^*$. Positive *n*'s are blue colored and negative *n*'s red. The inset of (b) illustrates how $B_n^*$ is determined in a density sweep.

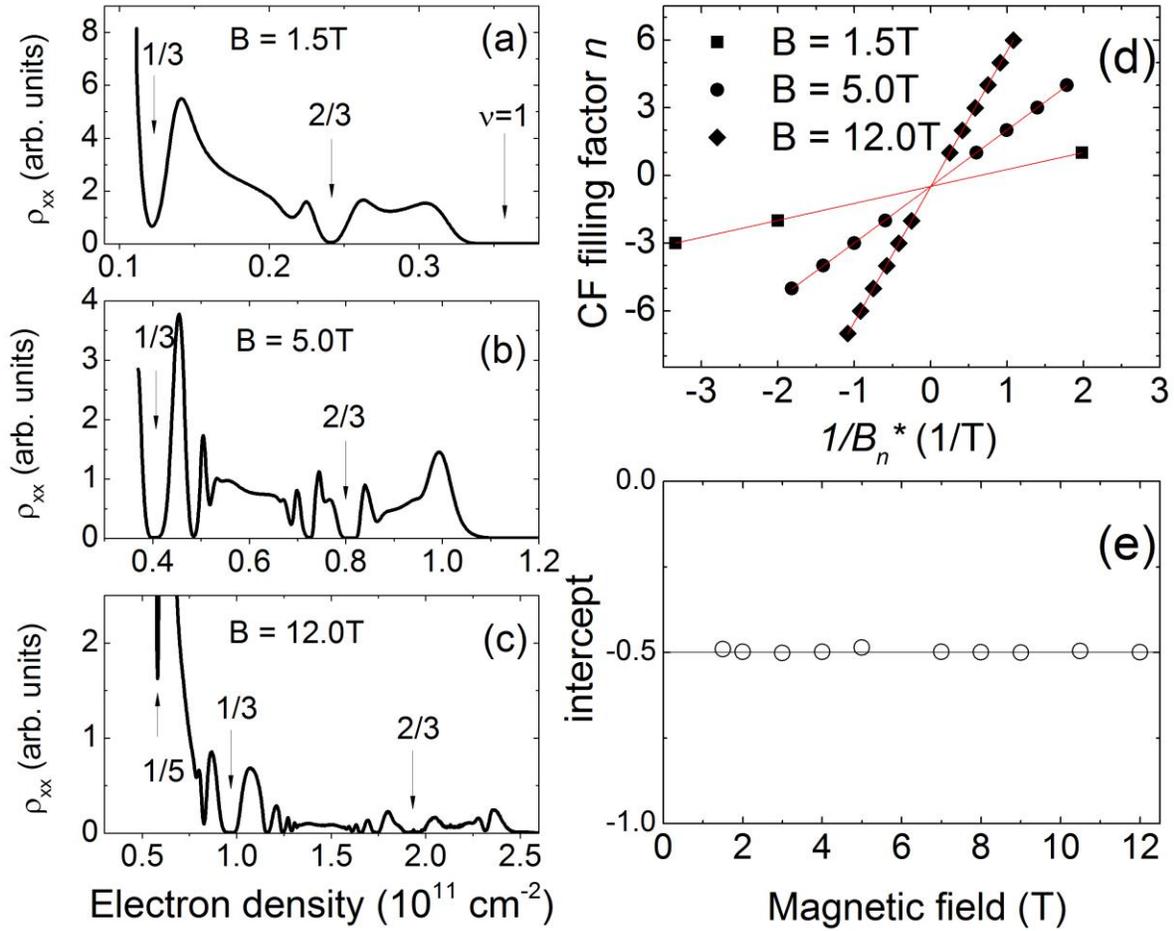

Figure 2: Shubnikov-de Haas oscillations around $\nu=1/2$ as a function of electron density ($n_e$, in units of $10^{11}$ $cm^{-2}$) in HIGFET-B at a few selective magnetic fields. B = 1.5T in (a), 5.0T in (b), and 12.0T in (c). The arrows mark the Landau level filling factors at $\nu=1/3$, 2/3, and 1, 1/5, respectively. (d) displays CF pseudo-Landau level filling factor $n$ vs $1/B_n^*$, where $B_n^* = B - 2n_eh/e$. (e) The bottom panel displays the intercept at $1/B_n^* = 0$ in (d) vs the external magnetic field.

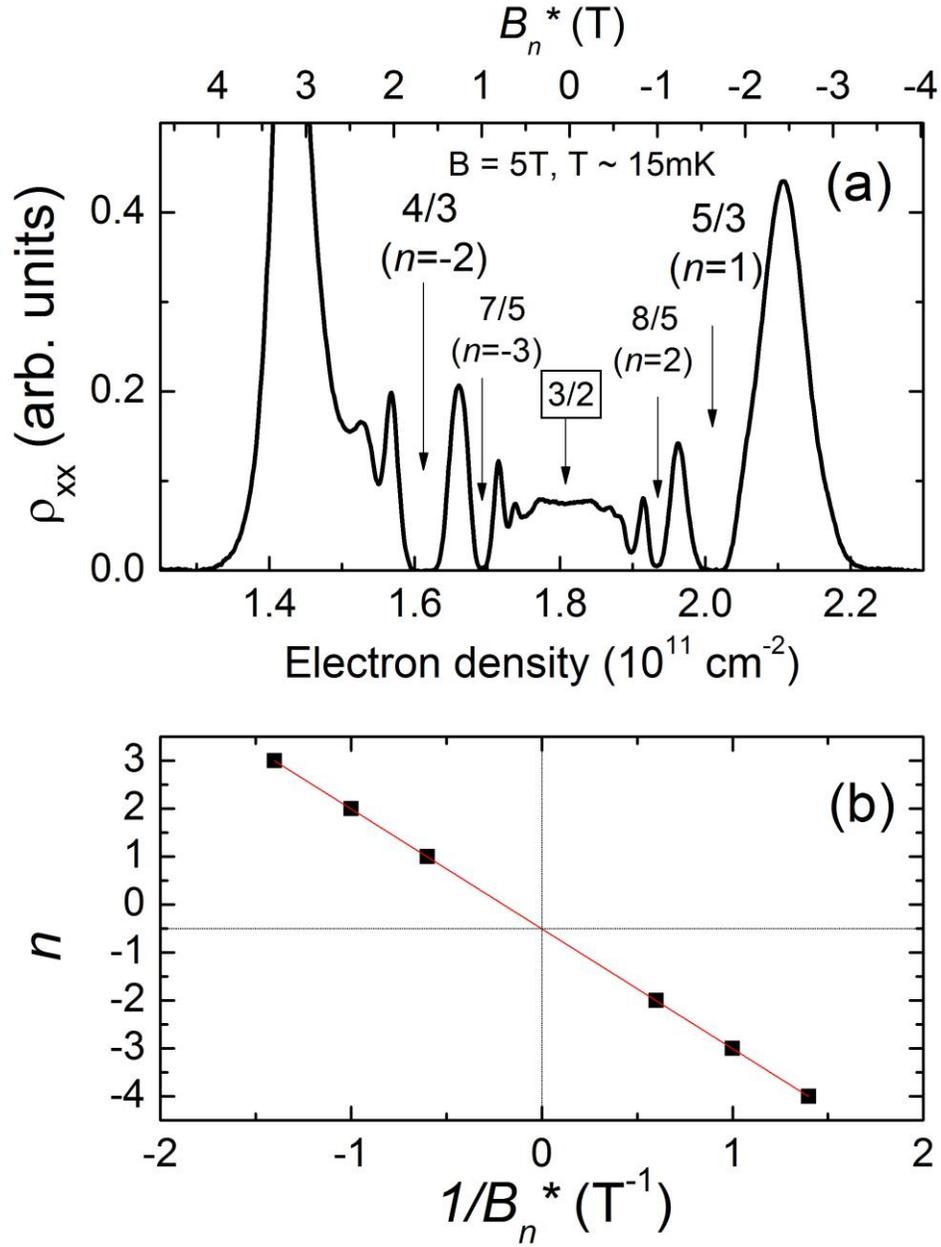

Figure 3: Composite fermion Shubnikov-de Haas oscillations around $\nu=3/2$ and Landau fan diagram. (a) shows $\rho_{xx}$ as a function of electron density ($n_e$, in units of $10^{11}$ $cm^{-2}$). (b) displays CF pseudo-Landau level filling factor $n$ vs $1/B_n^*$. The intercept at $1/B_n^* = 0$ is -1/2.

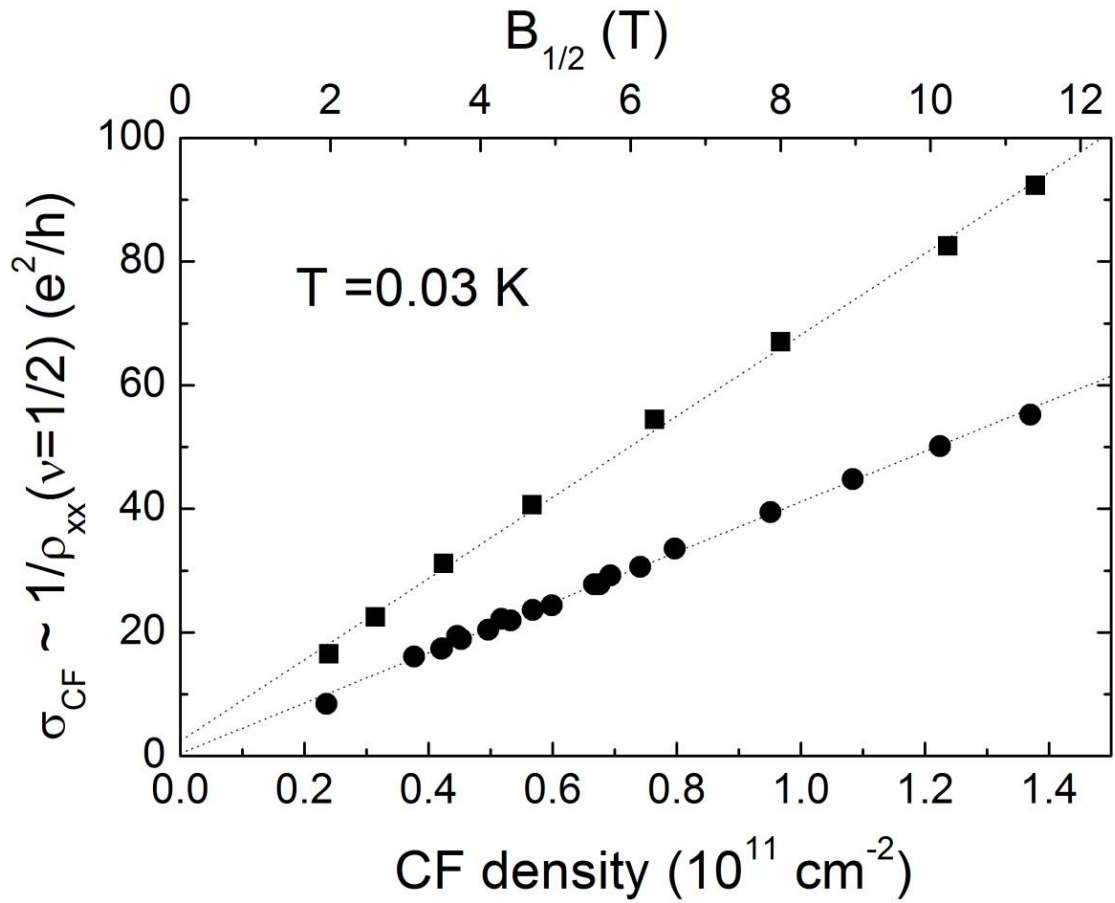

Figure 4: Conductivity of composite fermions at $\nu = 1/2$ in two HIGFETs. The two HIGFETs have different growth structures and, thus, peak mobilities. In the density range between $2\times10^{10}$ and $1.4\times10^{11}$ $cm^{-2}$, the conductivity displays linear density dependence in both samples.

# SUPPLEMENTARY INFORMATION

## Berry Phase and Anomalous Transport of the Composite Fermions at the Half-Filled Landau Level

W. Pan[1,*], W. Kang[2,*], K.W. Baldwin[3], K.W. West[3], L.N. Pfeiffer[3], and D.C. Tsui[3]

[1]Sandia National Laboratories, Albuquerque, NM 87185 USA

[2]Department of Physics, University of Chicago, Chicago, IL 60637 USA

[3]Department of Electrical Engineering, Princeton University, Princeton, NJ 08544 USA

In Fig. S1, we show the experimental observation of a self-similarity between the integer quantum Hall effect (IQHE) and the fractional quantum Hall effect (FQHE). In (a), we show the typical traces of magnetoresistance $R_{xx}$ and Hall resistance $R_{xy}$ as a function of magnetic field in a two-dimensional electron system. In (b), we shift the $R_{xx}$ curve horizontally so that the B field position at $\nu=1/2$ overlaps with $B = 0$. After this shift, it can be seen that the $\nu=1/3$ and 1 states, the 2/5 and 2 states, the 3/7 and 3 states *etc.* occur at the same B field positions, respectively. In (c), we shift the Hall trace diagonally so that the position of the 1/2 state overlaps with $B = 0$. Again after this shift, the one-to-one correspondence, 1/3 to 1, 2/5 to 2, 3/7 to 3, and 2/3 to -2, 3/5 to -3, 4/7 to -4 ..., between the IQHE and FQHE is observed. In other words, the change of Hall resistivity from half-filling, , $\Delta\rho_{xy} = (1/n)(h/e^2)$, determines the index n independently, in a way similar to determining the quantum number for integer quantum Hall states. Based on this observation, one can assign an effective integer filling factor to a FQHE state, i.e., filling factor 1 to the 1/3 state, 2 to 2/5, 3 to 3/, and -2 to 2/3, -3 to 3/5, -4 to 4/7, *etc.*

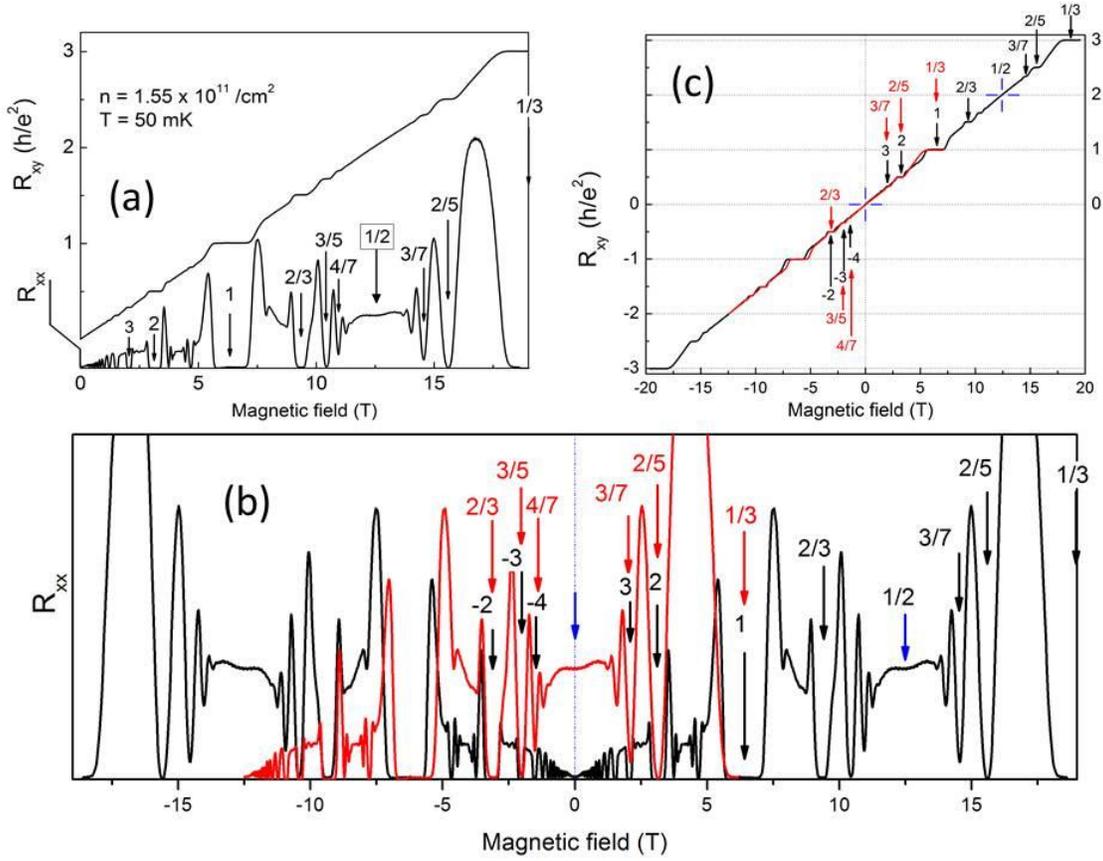

Fig. S1: Experimental demonstration of a self-similarity between the integer quantum Hall effect and the fractional quantum Hall effect.

To characterize the device quality of HIGFET, we use the Shubnikov-de Haas (SdH) measurements at low magnetic fields. In Figure S2 we show how the electron density was determined in HIGFET-B. Figure S2a shows the traces of $\rho_{xx}$ and $\rho_{xy}$ versus $V_g$ at B = 500mT in a semi-log plot. Figure S2b shows the traces of $\sigma_{xx}$ and $\sigma_{xy}$ versus $V_g$. Here, $\sigma_{xx} = \rho_{xx}/(\rho_{xx}^2 + \rho_{xy}^2)$ and $\sigma_{xy} = \rho_{xy}/(\rho_{xx}^2 + \rho_{xy}^2)$. The Landau level filling factor ν is assigned by the quantized Hall plateaus in the $\sigma_{xy}$ trace, $\sigma_{xy} = \nu \times e^2/h$. In the low $V_g$ regime, the strength of even and odd filling states is more or less equal. For large $V_g$, the strength of odd filling states is much weaker. In Figure S2c, we plot ν versus $V_g$. In Figure S2d, the electron density $n_e$ is plotted against $V_g$, where $n_e$ is in units of $10^{11}$ cm$^{-2}$ and $V_g$ in volt. $n_e$ is calculated by using $n_e=eB\nu/h$. A linear fit, $n_e = -0.154+1.184\times V_g$ is obtained. This linear relation is then used to convert $V_g$ into $n_e$ for all other traces at higher magnetic fields.

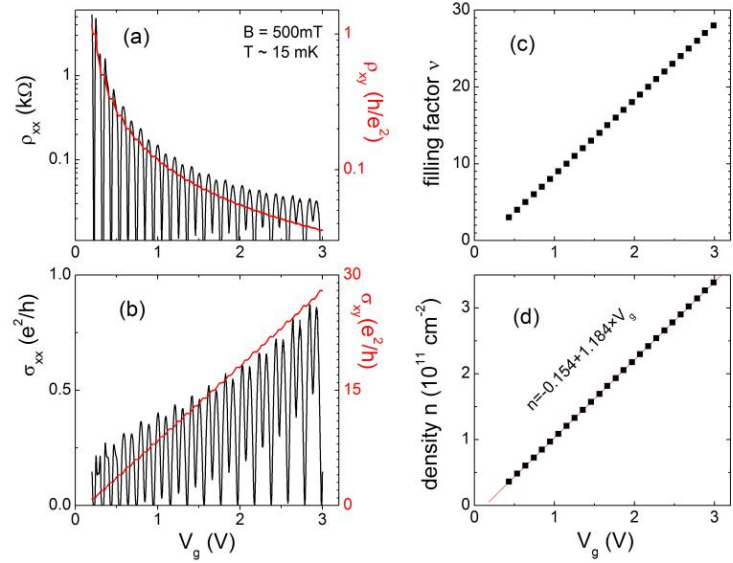

Figure S2a: $\rho_{xx}$ and $\rho_{xy}$ versus $V_g$ in a semi-log plot. The magnetic field is 500mT and the measurement temperature 15mK. Figure S1b shows $\sigma_{xx}$ and $\sigma_{xy}$ versus $V_g$ plot. Quantized Hall plateaus are clearly seen in the $\sigma_{xy}$ trace, from which the Landau level filling factor $\nu$ can be determined. Figure S1c shows the plot of Landau level filling factor $\nu$ versus $V_g$. Figure S1d shows the plot of electron density n versus $V_g$. The line is a linear fit.

In Fig. S3, we show the electron mobility as a function of density in HIGFET-B.

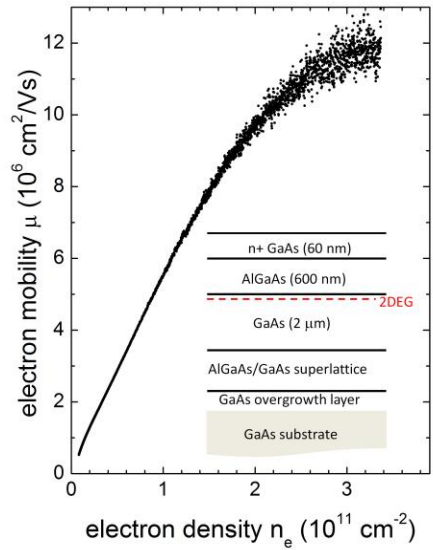

Fig. S3: Electron mobility versus density in HIGFET-B (need to get the real data from HIGFET-B). The measurement temperature is 15 mK. The inset shows the schematic growth structure of HIGFET.

The standard Shubnikov-de Haas oscillations were carried out in a high mobility specimen cut from the same GaAs quantum well wafer as the one used in Ref. S1. The quantum well thickness is 50 *nm*. An electron density of $n = 1.190 \times 10^{11}$ $cm^{-2}$ and mobility of $\mu \sim 13 \times 10^6$ $cm^2/Vs$ were achieved after a low temperature illumination with a red light-emitting-diode. Indium contacts, placed at the four corners and the middles of sample edges, were annealed at 420 °C for 8 minutes.

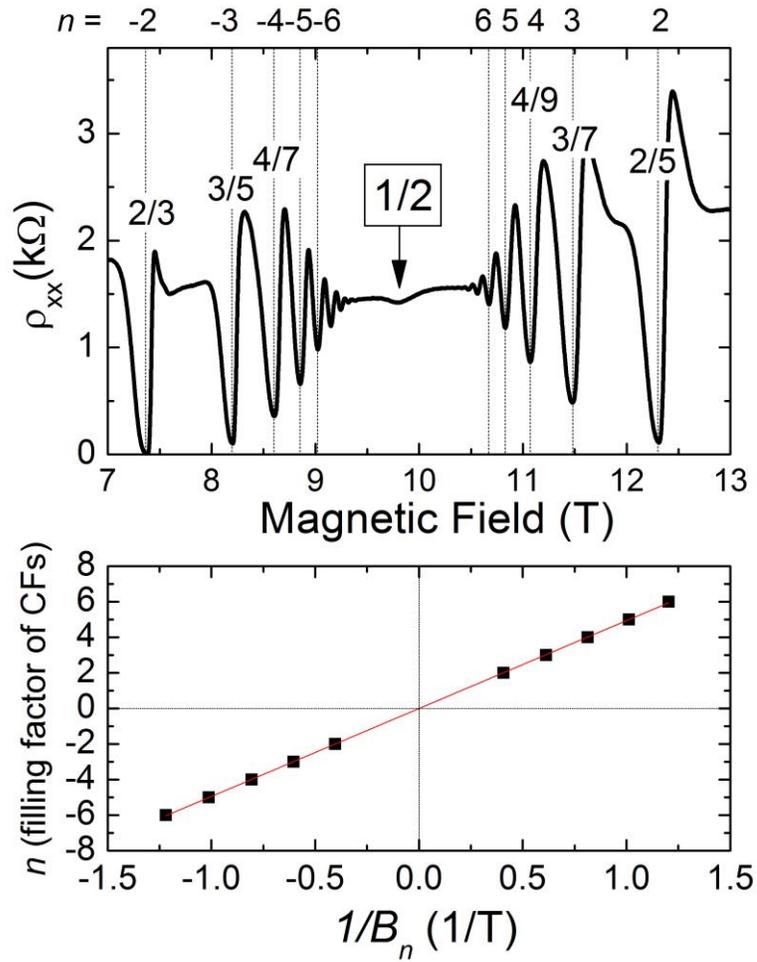

Figure S4: (a) shows the standard Shubnikov-de Haas oscillations (as a function of magnetic field at a fixed electron density) measured in a high quality 2DEG realized in a 50nm GaAs quantum well [S1]. The arrows mark the FQHE states at various Landau level fillings. The experimental temperature is 190 mK. (b) displays CF pseudo-Landau level filling factor *n* vs $1/B_n$. Here $B_n = B_\nu - B_{1/2}$. The intercept at $1/B_n = 0$ is 0.

Figure S4a shows the standard SdH oscillations measured in this high quality quantum well sample. The vertical lines mark the Landau level filling factors of electrons. The filling factors of CFs are also given for each FQHE state. The effective magnetic field of CFs is obtained by $B_n = B_\nu - B_{1/2}$. In Figure S4b, the CF Landau level filling factor n is plotted versus $1/B_n$. It is clearly seen that the intercept at $1/B_n = 0$ is zero for the 1/2 state in this analysis of standard SdH oscillations.

In the top panels of Fig. S5a and S5b, we show the density dependent SdH oscillations at the fixed magnetic fields of B = 3 and 7T, respectively. In their bottom panels, the CF Landau level filling factor $n$ is plotted versus $1/B_n^*$. The intercept at $1/B_n^* = 0$ for both B fields is very close to -0.5. In Fig. S5c, a summary of slope intercepts determined from the various density sweeps with different magnetic fields is shown. They all universally possess the value of -1/2, confirming the predicted Berry-phase of π for the CFs at ν = 3/2.

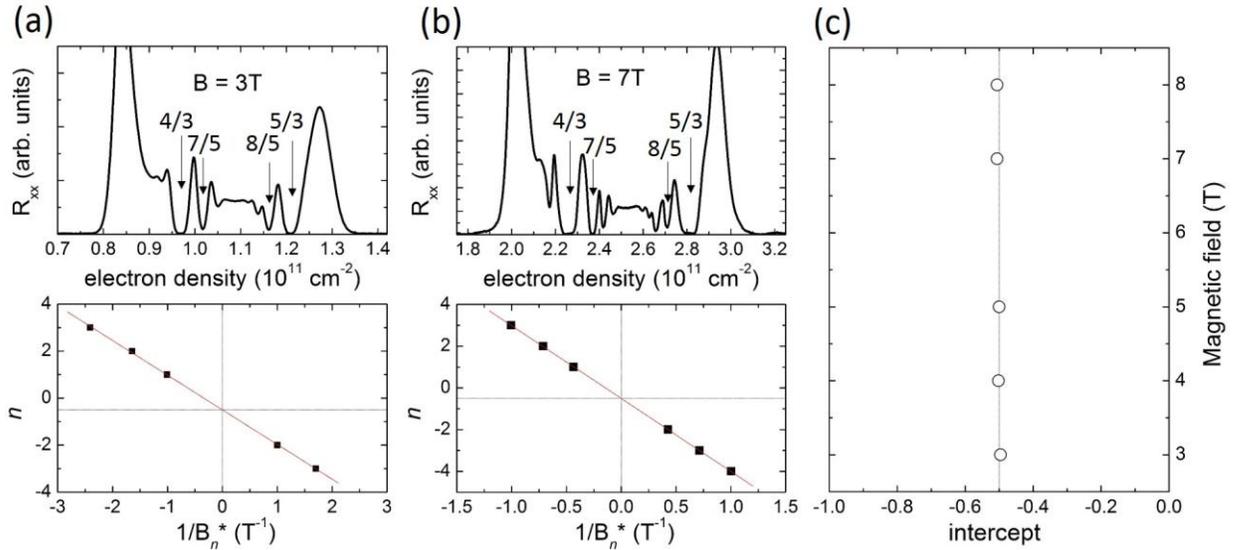

Fig. S5: the top panels of (a) and (b) show the Shubnikov-de Haas oscillations around ν=3/2 as a function of electron density ($n_e$, in units of $10^{11}$ $cm^{-2}$) in HIGFET-B at B = 3T (a) and 7T (b). The arrows mark the Landau level filling factors at ν=4/3, 7/5, 8/5, 5/3, respectively. The bottom panels of (a) and (b) display CF pseudo-Landau level filling factor n vs $1/B_n^*$, where $B_n^* = 3\times(B - 2n_e h/3e)$. (c) displays the slope intercept at $1/B_n^* = 0$ vs various external magnetic fields.

In Figure S6a we show the standard SdH oscillations around ν=3/2 [Ref. S2]. In Fig. S6b, the CF Landau level filling factor *n* is plotted versus $1/B_n$ for the standard SdH oscillations around ν=3/2. The intercept at $1/B_n = 0$ is -0.665, very close to the expected value of "-2/3".

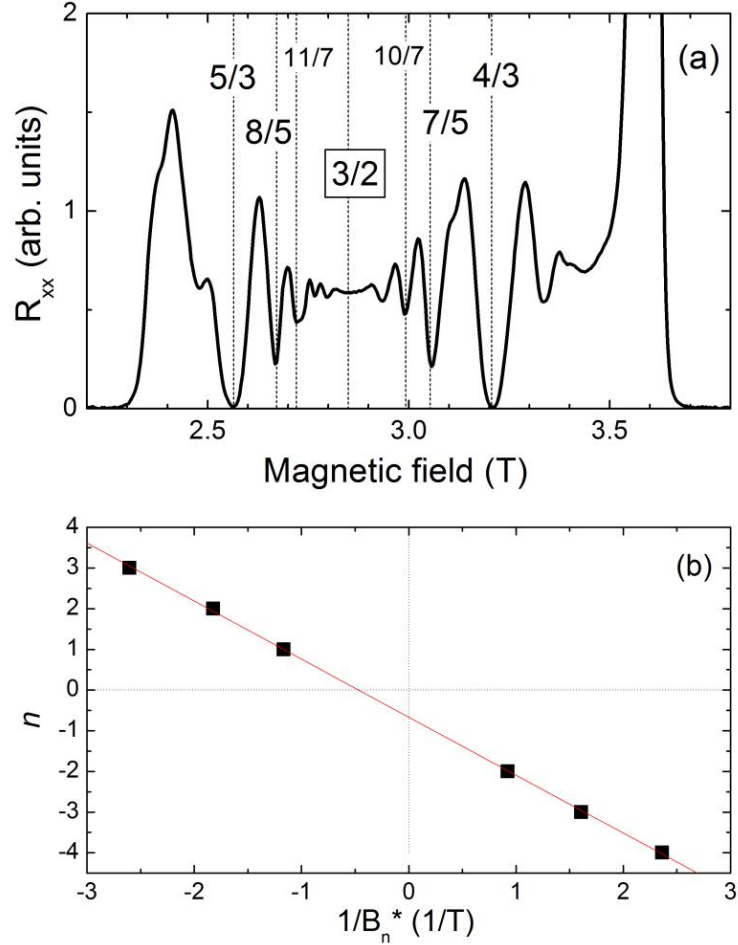

Figure S6: (a) shows the standard oscillations at a function of around ν=3/2. (b) displays CF pseudo-Landau level filling factor *n* vs $1/B_n$. Here $B_n = 3\times(B_v-B_{3/2})$. The intercept at $1/B_n= 0$ is -0.665, very close to -2/3.

In Fig. S7 we present the result of the density dependence of CF conductivity at ν=3/2 in HIGFET-B. The conductivity clearly shows a linear dependence over a large density range from ~ $1\times10^{10}$ to $1\times10^{11}$ $cm^{-2}$.

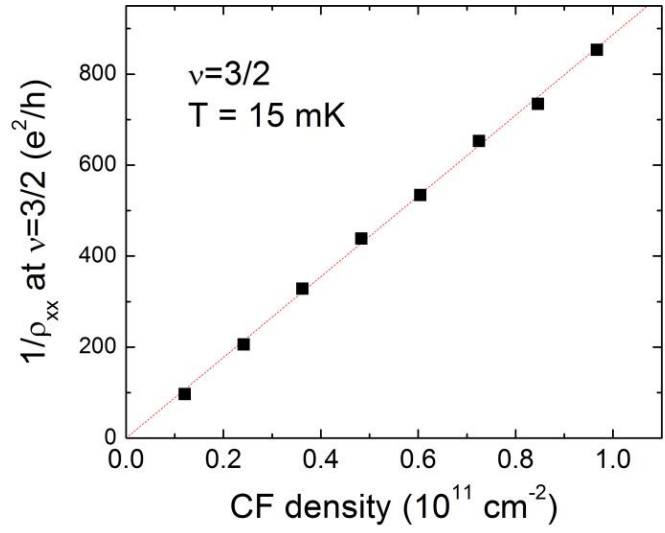

Figure S7: Conductivity ($\cong 1/\rho_{xx}$) of composite fermions at $\nu = 3/2$ in HIGFET-B. In the density range between ~$1\times10^{10}$ and ~$1\times10^{11}$ $cm^{-2}$, the conductivity displays linear density dependence.

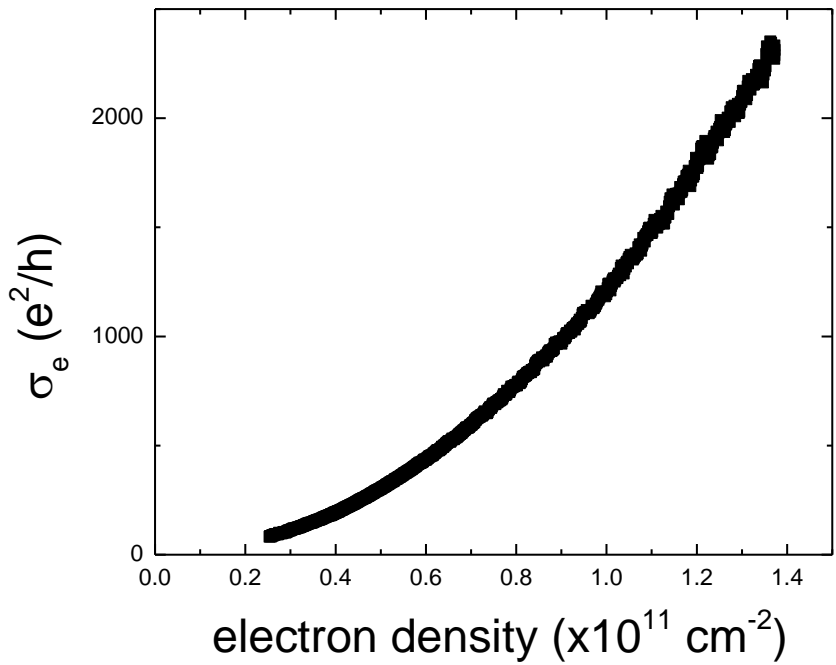

Figure S8: Density dependence of electron conductivity in HIGFET-A.

In Fig. S9, we deduce the quantum mobility and Landau level broadening parameter $\Gamma$ from analyzing the CF Shubnikov-de Haas (SdH) oscillations. Fig. S9a shows how the SdH oscillation amplitude is determined. The amplitude of SdH oscillations is then fitted to the equation of $\Delta R_{xx} = 4R_0 \times X/\sinh(X) \times \exp(-\pi/\mu_q B_n^*)$. Here $R_0$ is the value at $B_n^* = 0$, $X = 2\pi^2 k_B T/\hbar\omega_c$, $\omega_c = eB_n^*/m^*$ the CF cyclotron frequency, and $\mu_q$ the quantum mobility. In the B field range and at the temperature of 15 mK where the SdH oscillations were carried out, $X < 0.2$. Consequently, $X/\sinh(X) \approx 1-X^2/6 \approx 1$, within 1%. The above equation of $\Delta R_{xx}$ is then reduced to $\Delta R_{xx} = 4R_0 \times \exp(-\pi/\mu_q B_n^*)$. In Fig. S9b, $\ln(\Delta R_{xx})$ is plotted versus $1/B_n^*$. From the value of slope, $\mu_q$ is deduced. In Fig. S9c, $\mu_q$ is plotted as a function of CF density. Overall, $\mu_q \sim 1$ m$^2$/Vs and shows a weak CFs density dependence. The quantum time $\tau_q$ is calculated using $\tau_q = \mu_q m^*/e$. The CF effective mass is given by $m^*/m_0 = 0.26 \times \sqrt{B_{\nu=1/2}}$ [Ref. S3]. Here $B_{\nu=1/2}$ is the magnetic field at $\nu=1/2$, and $m_0$ the free electron mass. Finally, the Landau level broadening parameter $\Gamma = \hbar/\tau_q$ is plotted in Fig. S9c. In the density range between $0.5\times10^{11}$ and $1.4\times10^{11}$ cm$^{-2}$, $\Gamma$ is almost independent of CF density and $\Gamma \sim 0.15$ meV.

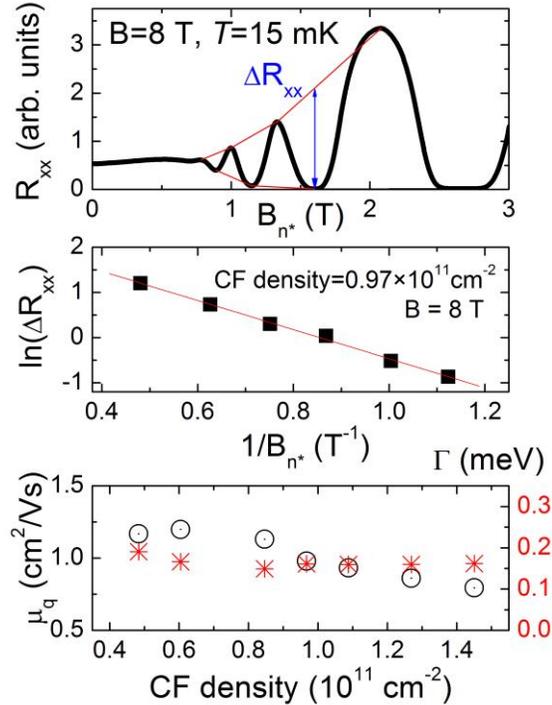

Fig. S9: (a) CF SdH oscillations as a function of effective magnetic field. The external field is fixed at B = 8T. (b) shows the Dingle plot. (c) $\mu_q$ (left y-axis) and $\Gamma$ (right y-axis) as a function of CF density.